\documentclass[aps,prl,twocolumn,showpacs,superscriptaddress, groupaddress]{revtex4-1}  
\usepackage{graphicx}  
\usepackage{dcolumn}   
\usepackage{bm}        
\usepackage{amssymb}   
\usepackage{amsmath}   
\usepackage{natbib}   
\usepackage{color}
\usepackage{CJK}
\begin{document}

\widetext

\title{Dispersive Readout of Room-Temperature Spin Qubits}
\vskip 0.25cm
\author{ J. Ebel}
\thanks{These two authors contributed equally}
\affiliation{TU M\"unchen, Walter Schottky Institut and Physik--Department, Am Coulombwall 4, 85748 Garching, Germany}
\affiliation{Munich Center for Quantum Science and Technology (MCQST), Schellingstr. 4, D-80799 M\"unchen, Germany}
\author{ T. Joas}
\thanks{These two authors contributed equally}
\affiliation{TU M\"unchen, Walter Schottky Institut and Physik--Department, Am Coulombwall 4, 85748 Garching, Germany}
\affiliation{Munich Center for Quantum Science and Technology (MCQST), Schellingstr. 4, D-80799 M\"unchen, Germany}
\author{M. Schalk}
\affiliation{TU M\"unchen, Walter Schottky Institut and Physik--Department, Am Coulombwall 4, 85748 Garching, Germany}
\affiliation{Munich Center for Quantum Science and Technology (MCQST), Schellingstr. 4, D-80799 M\"unchen, Germany}
\author{Andreas Angerer}
\affiliation{Vienna Center for Quantum Science and Technology, Atominstitut, TU Wien, Vienna, Austria.}
\author{J. Majer}
\affiliation{Vienna Center for Quantum Science and Technology, Atominstitut, TU Wien, Vienna, Austria.}
\affiliation{Shanghai Branch, CAS Center for Excellence and Synergetic Innovation Center in Quantum Information and Quantum Physics, University of Science and Technology of China, Shanghai 201315, China}
\affiliation{National Laboratory for Physical Sciences at Microscale and Department of Modern Physics, University of Science and Technology of China, Hefei 230026, China}
\author{F. Reinhard}
\email{friedemann.reinhard@wsi.tum.de}
\affiliation{TU M\"unchen, Walter Schottky Institut and Physik--Department, Am Coulombwall 4, 85748 Garching, Germany}
\affiliation{Munich Center for Quantum Science and Technology (MCQST), Schellingstr. 4, D-80799 M\"unchen, Germany}
\date{\today}

\begin{abstract}
We demonstrate dispersive readout of the spin of an ensemble of Nitrogen-Vacancy centers in a high-quality dielectric microwave resonator at room temperature. The spin state is inferred from the reflection phase of a microwave signal probing the resonator. Time-dependent tracking of the spin state is demonstrated, and is employed to measure the $T_1$ relaxation time of the spin ensemble. Dispersive readout provides a microwave interface to solid state spins, translating a spin signal into a microwave phase shift. We estimate that its sensitivity can outperform optical readout schemes, owing to the high accuracy achievable in a measurement of phase. The scheme is moreover applicable to optically inactive spin defects and it is non-destructive, which renders it insensitive to several systematic errors of optical readout and enables the use of quantum feedback. 
\end{abstract}

\pacs{{ 42.50 Pq, 76.30.Mi}} 
\maketitle
Ensembles of solid-state spin qubits, most prominently Nitrogen-Vacancy (NV) centers in diamond, are prominent candidates for a new generation of quantum sensors, promising sensitive magnetometers and gyroscopes in a compact device \cite{taylor08, ledbetter12, ajoy12, degen17}. The sensor signal is the expectation value of a spin component (typically $\langle\hat \sigma_z\rangle$), that needs to be measured to read out the sensor. 
For Nitrogen-Vacancy spin qubits, spin-dependent fluorescence provides a straightforward way to measure the spin, and has been the workhorse technique for readout in laboratory implementations of ensemble sensors \cite{wolf15}. This optical readout has also been employed in the first generation of integrated sensor devices \cite{stuerner19, kim19}, but it presents a roadblock to further integration, because miniaturisation of optics is difficult. The technique is also prone to systematic errors, such as a varying fluorescence background from spin-inactive NV centers in the neutral charge state NV$^0$. \par 
These limitations have prompted a search for all-electric readout techniques, that directly provide a measurement of the spin state as a current or voltage. 
Most prominently, spin-dependent photo-ionisation of NV centers has been used to induce a spin-dependent photocurrent \cite{bourgeois15, hrubesch17}, which has enabled photo-electric spin readout down to the level of a single spin \cite{siyushev19}. However, the readout accuracy of this all-electric method is limited by background impurities, dark currents, and fluctuations in electric properties, such as the ionisation cross section. \par
\begin{figure}
\includegraphics[ scale=1]{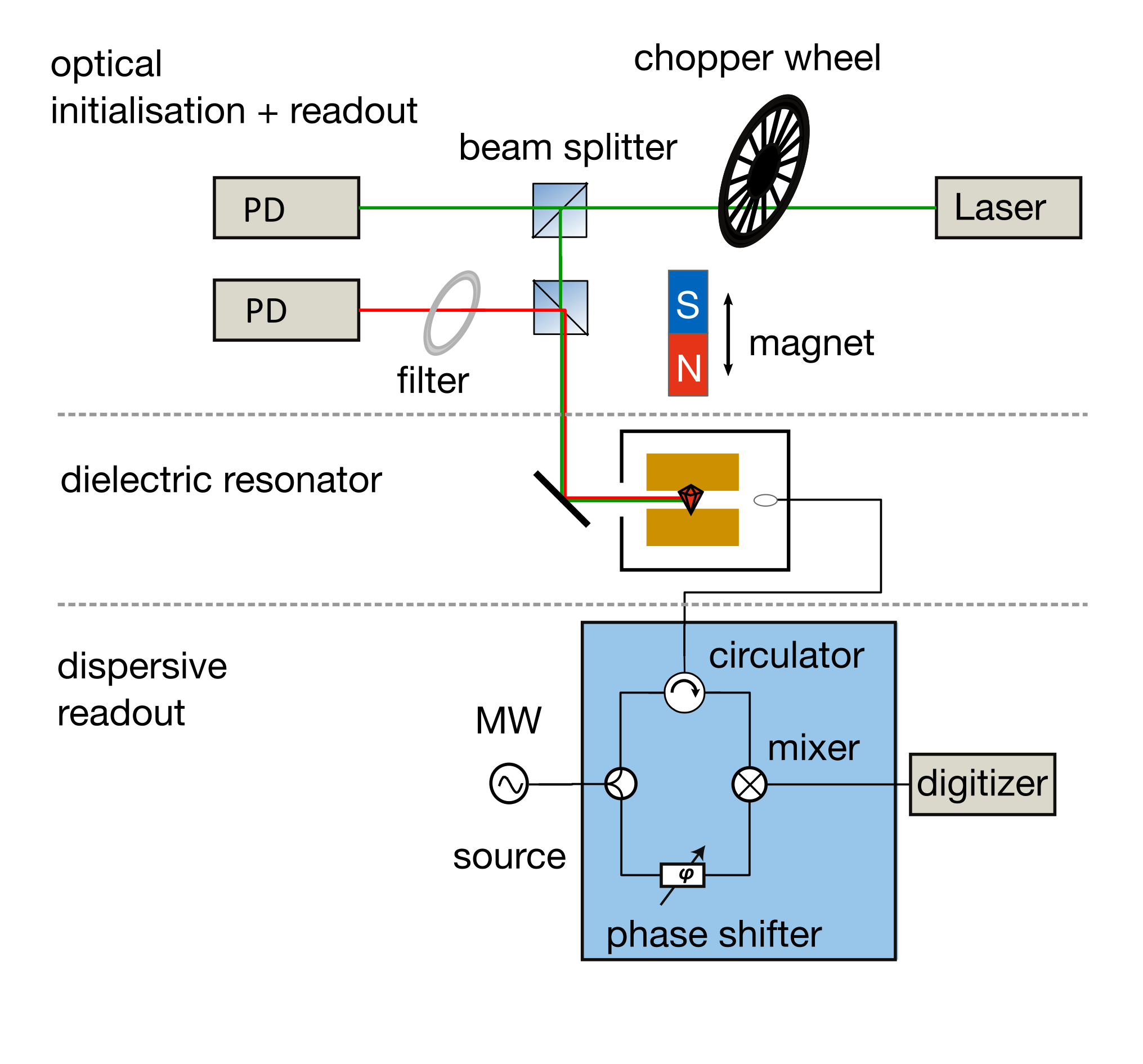}
\caption{\label{fig:fig1} Experimental setup. A densely NV-doped diamond is embedded in a stack of two cylindrical dielectric resonators with $Q\approx 6\cdot 10^3$. Laser excitation provides optical spin initialisation and optical readout. A microwave interferometer performs dispersive readout of the resonator. PD: photodiode; MW: microwave.}
\end{figure}
Experiments on spin ensembles at milli-Kelvin temperatures frequently employ dispersive spin readout in a superconducting resonator \cite{astner18}, where the spin signal is encoded in the phase of a microwave signal rather than a voltage or current. This technique is also the state-of-the art solution for single-shot readout of superconducting qubits \cite{wallraff05, vijay11}. Briefly, qubits ($\hat\sigma$) are coupled to a microwave cavity ($\hat a, \hat a^\dagger$), resulting in a system described by the Jaynes-Cummings Hamiltonian 
\begin{equation}
\hat H = \hbar \omega_c \hat a^\dagger \hat a + \hbar \omega_0 \hat \sigma_z + \frac{\hbar g}2 (\hat a \hat \sigma_+ + \hat a^\dagger \hat \sigma_-)
\label{eq:jaynes_cummings}
\end{equation}
with $\omega_0$ and $\omega_c$ denoting the qubit and cavity transition frequencies and $g$ the qubit-cavity coupling. Presence of the qubit imparts a spin-dependent dispersive shift to the cavity. To first order in $g/\Delta = g/(\omega_c - \omega_0)$, this shift is given by 
\begin{equation}
\delta_{\omega_c} = g^2/\Delta
\label{eq:dispersive_shift}
\end{equation} 
\cite{blais04}. It can be probed by measuring the transmission or reflection phase of a microwave resonant with the cavity. This phase varies linearly with $\Delta$ within a bandwidth of $\approx \omega_c/Q$ ($Q$ denoting the cavity quality factor) around to the cavity resonance $\omega_c + \delta_{\omega_c} $. Sensitive detection of a small shift $\delta_{\omega_c} $ hence requires a high $Q$. This is most easily achieved in superconducting cavities, which reach $Q$ values between $10^4$ (stripline transmission-line cavity \cite{blais04}) and $10^{10}$ (bulk-cavity \cite{kuhr07}).\par

Interestingly, comparably high ($10^4-10^5$) quality factors can be achieved at room temperature, in dielectric resonators made from low-loss high-permittivity ceramics. These devices have already been used for detection of electron paramagnetic resonance \cite{webb11}, based on measurements of absorption or dispersion, although their use is limited by background signals from intrinsic defects in the dielectric material \cite{friedlaender15}, overlapping with many relevant sample spins. This issue is of no concern for spins with an intrinsic zero-field splitting, such as the NV center. Dielectric resonators have already been interfaced with NV centers, to amplify driving pulses \cite{kapitanova18} and to provide resonant feedback in NV-based masers \cite{breeze18}. Masers have also been proposed as a magnetometry device \cite{jin15}, although these would be restricted to continuous-wave operation and would not provide a way for generic spin readout. 
\par
Here we demonstrate that dielectric resonators enable dispersive projective readout of the spin state in ensemble-based quantum sensors at room temperature. Our work has been performed in the setup displayed in Fig. \ref{fig:fig1}. We study a diamond densely doped with NV centres (created by electron irradiation and annealing of a 100 oriented type Ib diamond) interfaced to a cavity formed by a stack of two cylindrical dielectric resonators (diameter $16.8\;$mm, height $5.6\;$mm, $g/2\pi=2.4\cdot 10^{-2}\;$Hz estimated by an analytical model (\cite{kajfez86} ch. 4.4)). Stacking is employed to homogenise coupling to the cavity and to tune the resonance frequency close to the NV zero field splitting. This resonator is housed in a shielded enclosure and is probed by a microwave signal, magnetically coupled by a tuneable coupling loop. Probing is performed in a single-sided reflection geometry, where the phase of the reflected microwave arg$(S_{11})$ is measured by homodyne detection, and subsequently serves as the readout signal for the sensor. A strong laser ($532$nm, $300$mW) is employed to polarize the NV spins. This laser also implements optical spin readout as a complementary signal, recorded by monitoring fluorescence of the NV centers by a photodiode. A tuneable magnetic field is applied along the $001$ direction by a moveable permanent magnet. 
\par
\begin{figure}
\includegraphics[ scale=1]{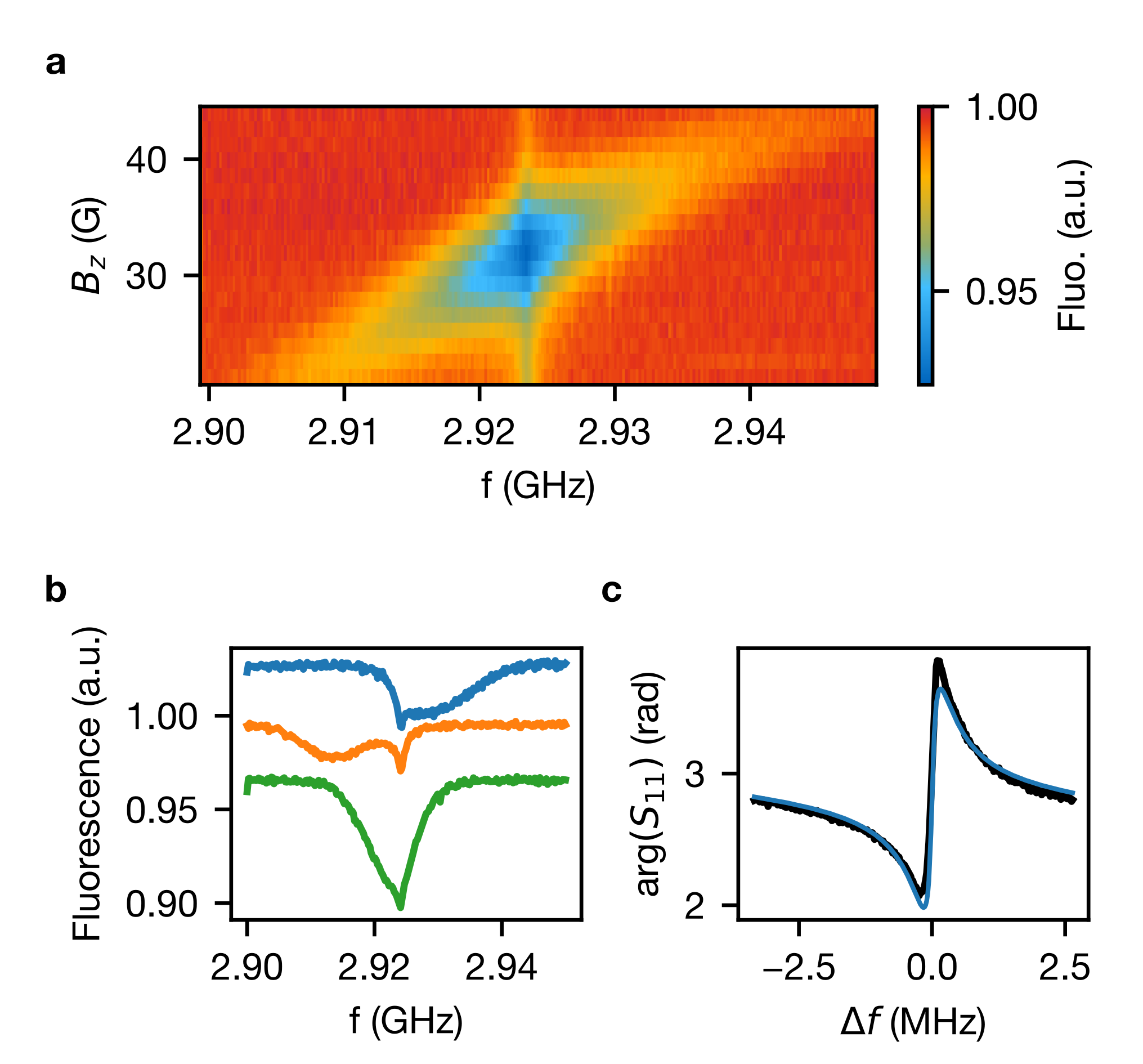}
\caption{\label{fig:fig2} Spin spectroscopy in the dielectric resonator.  a) optical spin readout for varying detuning. $\Delta$ is varied by changing the magnetic field, shifting the NV ensemble across the cavity resonance. An optical spin signal is evoked at the resonance frequencies $\omega_0$ and $\omega_c$ of the spin and the cavity. b) line plots of the optical signal for $B_z=26.5\;$G (orange), $B_z =32.5\;$G (green) and $B_z = 38.5\;$G (blue). Lines have been offset by increments of 0.03. c) Reflection phase of the resonator; black: measurement; blue: fit to the model of \cite{aitken76} with $Q= 6.0(1)\cdot 10^3$. }
\end{figure}
This setting allows for a study of NV-cavity coupling for a wide range of detuning $\Delta$ (Fig. \ref{fig:fig2}). For all values of $\Delta$, we observe that an optical spin signal (Figs. \ref{fig:fig2}a+b) is evoked at the two frequencies $\omega_0$ and $\omega_c$ of the spins and the cavity. At $\omega_0$, spins are resonantly driven, resulting in a significant signal despite inefficient off-resonant coupling into the cavity. At $\omega_c$, the drive is resonantly enhanced by the cavity, so that spins can be driven efficiently despite their detuning. We do not see a splitting of the cavity component when tuned into resonance with the spins, which indicates that the spin-cavity system is not in the strong-coupling regime. However, reaching this regime is not required for the demonstration of dispersive readout. \par
The reflection phase of the cavity varies steeply in vicinity of the resonance (Fig. \ref{fig:fig2}b). We fit this response to the model \cite{aitken76}
\begin{equation}
\text{arg}(S_{11}) = \frac{4\beta Q\Delta}{(2Q\Delta)^2 + (1-\beta^2)} + k\Delta +\phi_0 
\end{equation}
to obtain a quality factor of $Q=6.0(1)\cdot 10^{3}$ and a coupling coefficient of $\beta = 0.74(6)$. 
NV centers dispersively shift the cavity resonance (Fig. \ref{fig:fig2}c), so that arg$(S_{11})$ provides a direct electric measurement of the spin state for a microwave tuned to the cavity resonance. \par
\begin{figure}
\includegraphics[scale=1]{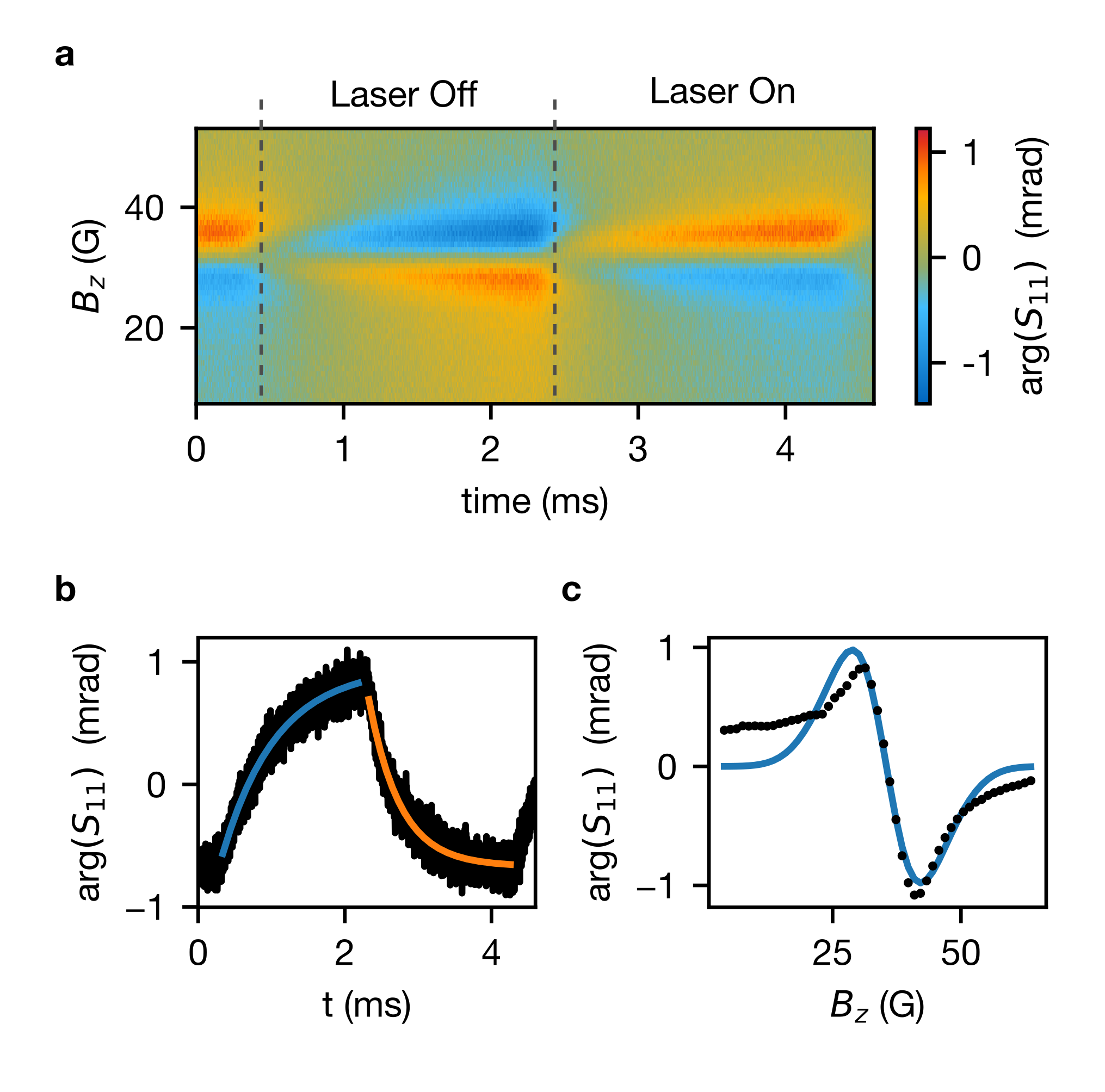}
\caption{\label{fig:fig3} Time-dependent dispersive spin readout. The polarisation laser is modulated so that spins alternate between optical polarisation (laser on) and decay by $T_1$ relaxation (laser off).  An offset has been subtracted from every line of data such that the temporal average is zero. a) temporal evolution of the reflection phase for varying $\Delta$. The dispersive signal grows for small values of $\Delta$ and changes sign with $\Delta$. b) temporal evolution for $B_z=32\;$G. Blue, orange: mono-exponential fits, yielding $T_{1,\text{Laser off}}=740(10)\;\mu$s and $T_{1,\text{Laser on}}=427(5)\;\mu$s; c) dispersive signal at maximum spin polarisation (average of a 200$\mu$s window around $t=2.2$ms). Blue: fit to spin-cavity model (see main text) with parameters $N_\text{NV}=2.0(1)\cdot 10^{12}$, $T_2^* = 18(1)\;$ns. Data deviates for magnetic fields of  $B<28\;$G, presumably due to a nonlinear field profile of the magnet in this range. This range has been excluded from the fit.}
\end{figure}
We employ this readout scheme for time-dependent tracking of the spin state (Fig. \ref{fig:fig3} a). Here, the laser is modulated by a mechanical chopper wheel, resulting in alternating bright and dark cycles of $2\;$ms duration. As spins polarize into their ground state under illumination, the dispersive phase shifts by up to $2\;$ mrad. In the dark phase of the cycle, $T_1$ relaxation resets the spins into a thermal state, resulting in a recovery of the dispersive phase shift to its original value. Both processes can be clearly recognized in the dispersive signal. A more thorough analysis (Fig. \ref{fig:fig3} b) reveals that both the buildup and decay of spin polarisation can be well described by mono-exponential decays, with a time constant of  $T_{1,\text{Laser off}}=740(10)\;\mu$s and $T_{1,\text{Laser on}}=427(5)\;\mu$s, respectively. The time constant is shorter for polarisation than for decay, which is likely explained by the high laser intensity required for optical readout. The time constant of $T_1$ relaxation is faster than values in comparable samples ($5\;$ms, \cite{jarmola12}). This is possibly due to the readout microwave, which remains present in the dark part of the cycle and could accelerate thermalisation. The dispersive shift grows for small detunings $\Delta$ and changes sign with $\Delta$, as expected from the expression $\delta_{\omega_c} = g^2/\Delta$. At the point of maximum spin contrast ($t=2.2\;$ms, Fig.\ref{fig:fig3}c), the data is well described by a numerical model computing the shift of Eq. \ref{eq:dispersive_shift} for an ensemble of NV centers with a Gaussian distribution of transition frequencies, the width of which is set by inhomogeneous broadening to $\sigma_\omega = 1/(2\pi T_2^*)$. \par

\begin{table}
\begin{tabular}{l|r}
Parameter & Value\\
\hline
$g$ & $2\pi \cdot 0.3\;$Hz\\ 
$\omega_0$ & $2 \pi \cdot 10^{10}\;$Hz\\ 
$Q$ factor & $10^4$\\
$\Delta$ & $2\pi \cdot 10^7\;$Hz\\
$T_2$ & $1\;$ms\\
$N$ & $10^{14}$
\end{tabular}
\caption{Parameters of an optimized device.}
\label{tab:scenario}
\end{table}

\begin{figure}
\includegraphics[scale=1]{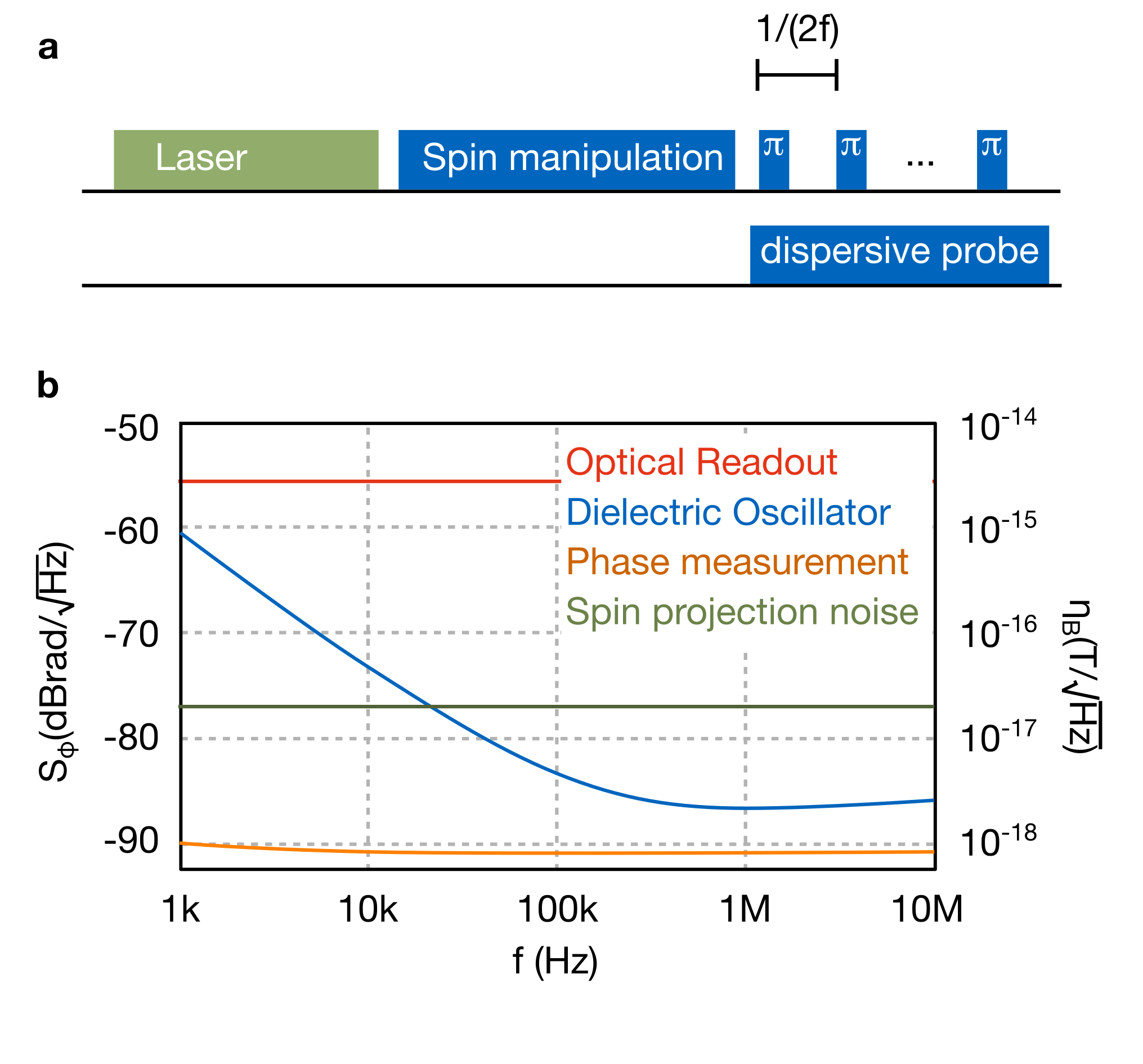}
\caption{\label{fig:fig4} Performance estimate for an optimized device. a) Readout sequence. Periodic flips during readout enable lock-in detection of the dispersive shift, removing phase noise below a cutoff frequency $f$. b) Performance estimate. Phase noise is estimated from published values of a dielectric oscillator (\cite{gronefeld17}, providing a conservative estimate) and a phase detector (\cite{rubiola99}, optimistic estimate). The performance of optical readout has been estimated by assuming a sensitivity 150 times worse than the spin shot noise limit, as in Ref. \cite{wolf15}.}
\end{figure}

We finally turn to a quantitative analysis of the ultimate performance that can be reached by the dispersive readout scheme. We assume the technical parameters of table \ref{tab:scenario}. The coupling strength is increased by an order of magnitude over the present work, which appears realistic by the use of a higher-frequency resonator with a smaller mode volume. The number of spins is increased by another two orders of magnitude to the level of reference \cite{breeze18}, so that the optimized cavity operates well within the strong-coupling regime ($\sqrt{g^2N} \gg \omega_c/Q, 1/T_2^*$). In a sensor, it will be desirable to choose the detuning large against all these parameters (e.g. to a value of $\Delta = 2\pi \cdot 10^7\;$Hz), in order to preserve sensitivity to magnetic fields. Still, the dispersive shift induced by the spins could be as large as 
\begin{equation}
\arg(S_{11}) = \frac{\pi Q N g^2}{\omega_0 \Delta} = 3\; \text{rad}.
\label{eq:dispersive_shift_optimized_device}
\end{equation}
The sensitivity of a sensor will be limited by the measurement accuracy on $\arg(S_{11})$, which will be limited by electronic phase noise and intrinsic noise of the dielectric resonator. Both of these mechanisms become stronger for decreasing frequencies, so that a naive implementation would be hampered by low-frequency noise such as thermal drift of the resonator. This problem can be overcome by lock-in schemes (Fig. \ref{fig:fig4} a), where the spin is modulated to create an oscillating signal. Such a modulation could be implemented by mere $T_1$ decay (as in our present work), or by a sequence of periodic control pulses. Readout sensitivity would then be limited by the phase noise $S_\phi(f)$ at the modulation frequency $f$. 
\begin{equation}
\eta_B = \frac{\hbar}{g\mu_B T_2}\frac{\omega_0 \Delta}{\pi Q g^2 N}S_\phi(f)
\label{eq:sensitivity}
\end{equation}
The phase noise $S_\phi (f)$ can be estimated from similar existing devices (Fig. \ref{fig:fig4}). Dielectric oscillators present a conservative estimate, because their phase noise  includes contributions from measurement, feedback, and intrinsic noise of the resonator. An optimistic estimate can be obtained from phase measurement circuits, which do not contain the latter two sources. In both cases, readout sensitivity can reach the limit of spin projection noise, which is currently out of reach for optical spin readout in ensembles. However, the worst-case scenario requires modulation at a frequency of $100\;$kHz or higher. We note that the microwave power required for a measurement at the spin shot noise limit would be still small much less than the Rabi frequency required to drive spin flips. Readout would require a minimum flux of $N/T_2$ microwave photons, populating the resonator with an average $2\pi N Q/ [\omega_c T_2]$ photons, corresponding to a Rabi frequency of $\Omega_R \approx 2\pi \cdot 200\;$Hz.  \par
In summary, we have demonstrated a dispersive approach to spin readout in quantum sensors. While its sensitivity promises to outperform established techniques, its implementation will not be without challenges. In particular, dispersive readout requires a narrowband cavity, where fast pulsed control is not easily implemented. Combining both requirements would be most straightforward in a dual-port cavity with the ability of rapid Q-switching \cite{rinard02}. \par
The technique has several complementary advantages over optical readout beyond sensitivity. It is applicable to a wider range of spin species than optical readout, as there are spins that can be optically polarised, but cannot be efficiently read out optically (such as the Silicon-Vacancy in Silicon-Carbide \cite{soltamov12}). Cavity readout is moreover non-destructive, so that spin states can be weakly probed during their evolution, as demonstrated here for the measurement of spin relaxation times. This is a crucial requirement for the implementation of quantum feedback schemes \cite{wiseman93, vijay12}. It also mitigates several problems of optical readout, such as a background of luminescent but spin-inactive centers in the neutral charge state NV$^0$. These would contribute fluorescence background but no dispersive signal. Finally, the mere technical simplicity of the scheme will enable straightforward integration into compact devices as it is required for large-scale application of quantum technologies. 
\par
This work has been supported by BMBF (Quant-ERA project MICROSENS), FWF (Quant-ERA SUMO), the European Union (Horizon 2020 research and innovation programme, grant agreement No 820394 (ASTERIQS)),  as well as by the Deutsche Forschungsgemeinschaft (DFG, German Research Foundation) under the German Excellence Strategy -- EXC-2111 -- 390814868 and Emmy Noether grant RE3606/1-1. During compilation of the manuscript, we became aware of related work \cite{eisenach20}.

\bibliography{dispersive_readout_spin_qubits,revtex-custom}

%
%


\end{document}